\documentclass[aps,superscriptaddress,eqsecnum,nofootinbib,showpacs,preprintnumbers]{revtex4}
\usepackage{graphicx,epsfig}
\usepackage{amsmath}
\usepackage {amssymb}
\usepackage[section]{placeins}

\newcommand{\be}{\begin{eqnarray}}
\newcommand{\ee}{\end{eqnarray}}
\newcommand{\bea}{\begin{eqnarray}}
\newcommand{\nn}{\nonumber}
\newcommand{\eea}{\end{eqnarray}}

\def\m{\mu}
\def\n{\nu}

\def\s{\sigma}

\def\m{\mu}
\def\n{\nu}

\def\s{\sigma}

\newcommand{\fr}{\frac}

\begin{document}

\title{Superradiant Instabilities in a Class of Scalar-Tensor Horndeski Theory}

\author{Theodoros Kolyvaris }
\email{teokolyv@mail.ntua.gr} \affiliation{Instituto de
F\'{i}sica, Pontificia Universidad Cat\'olica de Valpara\'{i}so,
Casilla 4059, Valpara\'{i}so, Chile.}

\author{Marina Koukouvaou}
\email{makou.5316@gmail.com} \affiliation{Physics Division,
National Technical University of Athens, 15780 Zografou Campus,
Athens, Greece.}

\author{Antri Machattou}
\email{andrimachattou@hotmail.com} \affiliation{Physics Division,
National Technical University of Athens, 15780 Zografou Campus,
Athens, Greece.}

\author{Eleftherios Papantonopoulos}
\email{lpapa@central.ntua.gr} \affiliation{Physics Division,
National Technical University of Athens, 15780 Zografou Campus,
Athens, Greece.}

\date{\today}

\begin{abstract}

We study the superradiance effect in a class of scalar-tensor Horndeski theory.  We first study the dynamics of a massive charged scalar wave scattered off the horizon of a Reissner-Nordstr\"om black hole which except its canonical coupling to gravity it is also coupled kinetically  to curvature. We find that a trapping potential is formed outside the horizon of a Reissner-Nordstr\"om black hole, due to this coupling,   and as the strength of the new coupling  is increased, the scattered wave   is  superradiantly amplified, resulting to the instability of the Reissner-Nordstr\"om spacetime. We then considered the backreacting effect of the scalar field coupled to curvature interacting with the background metric and we study the superradiant effect and find the superradiance conditions of a massive charged wave scattered off the horizon of a Horndeski black hole.

\end{abstract}

\maketitle

\section{Introduction}

It was shown by Penrose \cite{Penrose:1971uk} that if a particle accretes  into a Kerr black hole then this process may result to the extraction of energy from the black hole.
 This process suggests that energy is extracted
from the black hole. The same process can be realized if you scatter waves off the black hole. It was found by Misner \cite{Misner:1969hg} that this happens if
the relation
\be \omega < m \Omega \label{rad1} \ee
is satisfied, where $
\omega$ is the frequency of the incitant wave and $\Omega$ is the rotational frequency of the black hole. Later it was shown by
Teukolsky \cite{Teukolsky:1973ha} that this
applification mechanism also works in the case of electromagnetic and gravitational waves if the condition (\ref{rad1}) is satisfied.
Further Bekenstein showed \cite{Bekenstein} that the Misner process can also be realized by
extracting charge and electrical energy from a
charged black hole. He showed that this is a  consequence of the
Hawking's theorem that the surface area of a black
hole cannot decrease.

Introducing a reflecting mirror it was suggested  by   Press and Teukolsky \cite{PressTeu1} that the wave will be amplified
as a result of the
bounce back and forth between the black hole and the mirror. In the case of a rotating black hole, the role of the reflecting mirror can be played by the mass of the scalar field because the  gravitational force binds the massive field and
keeps it from escaping to infinity once the condition    (\ref{rad1}) is satisfied.
 As a consequence, the rotational energy
extracted from the black hole by the incident field grows
exponentially over time. A detailed investigation of the superradiant amplification of a wave scattered off a Kerr black hole surrounded by a mirror,  was carried out in \cite{Cardoso:2004nk}.
In the case of extracting charge and electric energy from a  Reissner-Nordstr\"om black hole, the Misner condition (\ref{rad1}) is modified \cite{Bekenstein}  to
\begin{equation}\label{rad2}
\omega<q\Phi ~,
\end{equation}
where $q$ is the charge coupling constant of the field and
$\Phi$ is the electric potential of the charged black hole.
The superradiant scattering
of charged scalar waves in the regime (\ref{rad2}) may lead to an
instability of the Reissner-Nordstr\"om spacetime. Recently there is a study of the superradiance effect of a black hole immersed in an expanding Universe \cite{Kiorpelidi:2018the}. Sufficient evidence was found that there is extraction of energy from the neutral McVittie black hole because of the shrinking of its apparent horizon as the Universe expands.

The stability of Reissner-Nordstr\"om  black holes under  neutral gravitational
and electromagnetic perturbations was established  by
Moncrief \cite{Monf1,Monf2}. Evidence was provided in \cite{Hod:2013eea,Hod:2013nn,Hod:2015hza,Hod:2016kpm}  for the
stability of charged Reissner-Nordstr\"om  black holes  under charged scalar
perturbations.  The stability of
extremal braneworld charged  holes was studied in \cite{Zhang:2013haa} were it was shown that if the spacetime dimension is higher than four,
the superradiant amplification can occur.
 Also  the Einstein-Maxwell-Klein-Gordon equations
for a spherically symmetric scalar field
scattering off a Reissner-Nordstr\"om  black hole in asymptotically flat spacetime were considered in \cite{Baake:2016oku} and a superradiant instability was found. The superradiance instability of  charged black holes placed in a cavity was studied in \cite{Herdeiro:2013pia,Degollado:2013bha,Sanchis-Gual:2015lje,Sanchis-Gual:2016tcm} (for a recent review on superradiance see \cite{Brito:2015oca}).

The recent developments in AdS/CFT correspondence \cite{Maldacena:1997re} introduces an AdS spacetime  as a natural reflecting boundary on which the
reflecting wave keeps amplified  driving the system unstable. This results in the formation of hairy  black holes with a charged scalar field trapped outside the black hole in which the electric repulsion balances the scalar condensate against gravitational collapse. Hairy black holes were constructed in  global AdS$_5$ spacetime in \cite{Basu:2010uz,Bhattacharyya:2010yg,Dias:2011tj,Basu:2016mol}. The superradiant amplification of a wave packet may result to the destabilization of the AdS spacetime itself \cite{Cardoso:2004hs,Cardoso:2013pza,Bosch:2016vcp,Dias:2016pma}. If we perturb the AdS spacetime with a scalar field  the system evolves towards the formation of a black hole \cite{Bizon:2011gg,Dias:2012tq,Buchel:2012uh,Bizon:2015pfa}. The stability of near  extremal and extremal  charged hairy black hole solution under charged massive scalar field perturbations was studied in \cite{Gonzalez:2017shu}.

The application of the  AdS/CFT correspondence to condensed matter
systems (for a review see \cite{Hartnoll:2009sz}) had revived the
interest on the dynamics of a scalar field outside a black hole
horizon and its stability. The transition of metalic state to a superconducting
state which is a strongly-coupled problem in condensed matter
physics can be described by its dual weekly-coupled gravity
problem using the AdS/CFT correspondence \cite{Maldacena:1997re}.
It was shown that the effective mass of the
scalar field which is trapped
just outside the horizon of a charged black hole \cite{Gubser:2005ih,Gubser:2008px} becomes negative  breaking in this
way an Abelian gauge symmetry outside the horizon of the
Reissner-Nordstr\"om black hole resulting to an instability of the Reissner-Nordstr\"om spacetime.

In the case of de-Sitter  charged black holes instabilities were found  in higher dimensions. It was showed in \cite{Konoplya:2008au,Konoplya:2013sba}
 that higher dimensional Reissner-Nordstr\"om-de Sitter black holes are gravitationally unstable for large values of the electric charge in $D≥7$ spacetime dimensions. The existence of
such instability was proved analytically in the near-extremal limit \cite{Cardoso:2010rz}. In four dimensions
a superradiance instability was found \cite{Zhu:2014sya} in the  Reissner-Nordstr\"om-de Sitter black holes
against charged scalar perturbations with vanishing angular momentum, l = 0.

In this work we study the effect of charged scalar perturbations on the stability of local solutions of  theories in which a scalar field is kinetically coupled to curvature.
These theories belong to a general class of
scalar-tensor gravity theories resulting from the Horndeski
Lagrangian \cite{Horndeski:1974wa} which
recently rediscovered \cite{Deffayet:2011gz},
 give second-order field equations and contain as a subset a theory
which preserves classical Galilean symmetry \cite{Nicolis:2008in,
Deffayet:2009wt, Deffayet:2009mn}.

This derivative coupling because it  has the dimensions of length squared,  redefines other possible scales present in the theory like the cosmological constant, giving in this way various black hole solutions \cite{Kolyvaris:2011fk,Rinaldi:2012vy,Kolyvaris:2013zfa,Babichev:2013cya},
while if one considers the gravitational collapse of a scalar field coupled to the Einstein tensor then a  black hole is formed
 \cite{Koutsoumbas:2015ekk}.
The presence of the derivative coupling has different behaviours during the cosmological evolution. It acts as a friction term in the inflationary period \cite{Amendola:1993uh,Sushkov:2009hk,germani} and also it gives self-acceleration or self-tuning cosmological solutions \cite{Linder:2013zoa,Babichev:2016kdt}. Moreover, it was found that at the end of
inflation in the preheating period, there is a suppression of heavy particle production  as the derivative coupling is increased. This was attributed to the fast decrease of  kinetic
energy of the scalar field due to its  wild oscillations \cite{Koutsoumbas:2013boa}. This change of the kinetic energy of the scalar field to Einstein tensor allowed to holographically simulate the effects of a high concentration of impurities in a material \cite{Kuang:2016edj}.

The above discussion indicates that one of the main effects of the kinetic coupling of a scalar field to Einstein tensor
is that  gravity  influences strongly  the  propagation of the scalar field compared to a scalar field minimally coupled to gravity.
We will use this property of the charged scalar field coupled to Einstein tensor to study its behaviour outside the horizon of charged black holes. First we will study the case of a test scalar field coupled to Einstein tensor scattered off the horizon of a  Reissner-Nordstr\"om black hole.   Will  show
that this new dimensionful
derivative coupling  provides a scale for a confining potential, which is an effect
similar to the AdS radius provided by the cosmological constant, and in the same time modifies the Bekenstein's superradiance condition (\ref{rad2}) with the derivative coupling appearing explicitly in the superradiance condition. Then for a wide range of parameters satisfying the supperadiant condition, the charged scalar field will be trapped in this confining potential leading to a superradiant instability of the Reissner-Nordstr\"om black hole. The stability of  Reissner-Nordstr\"om and Kerr spacetimes   was discussed in \cite{Chen:2010qf,Chen:2010ru,Ding:2010fh} in a different context, calculating the quasinormal frequencies and  the greybody factors  in the presence of the derivative coupling.

These results of the superradiance instabilities of the Reissner-Nordstr\"om black hole indicate that the background black hole may acquire scalar hair \cite{Dias:2016pma} and therefore we will study the backreacted effect. We will allow the scalar field coupled to Einstein tensor to backreact to a  charged spherical symmetric background. This will lead to the generation of hairy charge Galileon black holes. In these solutions  the derivative coupling appears as a parameter in these hairy solutions and then we will study how the superradiance effect and the superradiance conditions will be modified in the presence of this derivative coupling in the cases of a time dependent scalar hair  \cite{Babichev:2015rva} and of a static one \cite{Cisterna:2014nua}.

The  work is organized as follows. In Section \ref{instability} we will review the stability of a Reissner-Nordstr\"om black hole under charge scalar perturbations of a scalar field minimally coupled to gravity. In Section \ref{dercoupl} we introduce the derivative coupling of a scalar field to the Einstein tensor and we discuss the stability of the background Reissner-Nordstr\"om black hole under charge scalar field perturbations. In Section \ref{Galileon1} we study the superradiance effect of  the charged Galileon black hole with time-dependent scalar hair, in  Section \ref{Galileon2}  the superradiance in the charged Galileon black hole with static hair while in Section \ref{conclusion} are our conclusions.

\section{Supperradiant Stability of  the Reissner-Nordstr\"om black hole}
\label{instability}

In this Section we will review the superradiant stability of the Reissner-Nordstr\"om black hole discussed in \cite{Hod:2013eea,Hod:2013nn}. Consider a Reissner-Nordstr\"om black hole of mass $M$ and
electric charge $Q$ with a metric
\begin{equation}\label{Eq3}
ds^2=-f(r)dt^2+{1\over{f(r)}}dr^2+r^2(d\theta^2+\sin^2\theta
d\phi^2)\ ,
\end{equation}
where
\begin{equation}\label{Eq4}
f(r)\equiv 1-{{2M}\over{r}}+{{Q^2}\over{r^2}}\  .
\end{equation}

A massive charged scalar field was considered in the Reissner-Nordstr\"om spacetime and its dynamics was described by the
 the Klein-Gordon equation
\begin{equation}\label{Eq5}
[(\nabla^\nu-iqA^\nu)(\nabla_{\nu}-iqA_{\nu}) -\mu^2]\Psi=0\  ,
\end{equation}
where $A_{\nu}=-\delta_{\nu}^{0}{Q/r}$ is the electromagnetic
potential of the black hole. Here $q$ and $\mu$ are the charge and
mass of the field, respectively. Consider a scalar wave of the form
\begin{equation}\label{Eq6}
\Psi_{lm}(t,r,\theta,\phi)=e^{-i\omega t} R_{lm}(r) Y(\Omega) \,,
\end{equation}
where the angular part is given by
\begin{equation}
Y(\Omega)=e^{im\phi}S_{lm}(\theta)  \,,\label{wave}
\end{equation}
where $\omega$ is the conserved frequency of the mode, $l$ is the
spherical harmonic index, and $m$ is the azimuthal harmonic index
with $-l\leq m\leq l$. The frequency mode $ \omega $ is in general complex consisting of real part $\omega_R$ and an imaginary part $\omega_I$. Then, if  $\omega_I>0$ we expect a growing mode to develop, resulting to an  instability
\cite{Kokkotas:1999bd,Nollert:1999ji,Konoplya:2011qq}.  In the corresponding holographic picture of a fluid undergoing a phase transition, $\omega_I$ defines the relaxation time $T=1/\omega_I$ which is  needed for this instability to reach thermal equilibrium  \cite{Horowitz:1999jd}.

 In the above decomposition $R$ and $S$ denote the
radial and angular part and the radial Klein-Gordon equation is given by \cite{Hod:2013nn}
\begin{equation}\label{Eq7}
\Delta{{d} \over{dr}}\Big(\Delta{{dR}\over{dr}}\Big)+UR=0\ ,
\end{equation}
where
\begin{equation}\label{Eq8}
\Delta\equiv r^2-2Mr+Q^2\  ,
\end{equation}
and
\begin{equation}\label{Eq9}
U\equiv(\omega r^2-qQr)^2 -\Delta[\mu^2r^2+l(l+1)]\  .
\end{equation}

To solve the radial equation (\ref{Eq7}) we impose boundary conditions of purely ingoing waves at the
black-hole horizon and a decaying  solution at spatial infinity
\begin{equation}\label{Eq11}
R \sim e^{-i (\omega-qQ/r_H)y}\ \ \text{ as }\ r\rightarrow r_H\ \
(y\rightarrow -\infty)\ ,
\end{equation}
and
\begin{equation}\label{Eq12}
R \sim y^{-iqQ}e^{-\sqrt{\mu^2-\omega^2}y}\ \ \text{ as }\
r\rightarrow\infty\ \ (y\rightarrow \infty)\  ,
\end{equation}
where the ``tortoise" radial coordinate $y$ is defined by
$dy=(r^2/\Delta)dr$. Because the energy of the incident wave is positive, the frequencies are restricted to the  superradiant regime
(\ref{rad2}), and the boundary condition (\ref{Eq11}) describes an
outgoing flux of energy and charge from the charged black hole
\cite{Bekenstein,PressTeu1}. As it was shown in \cite{Damour:1976kh} these boundary conditions
lead to  a
discrete set of resonances $\{\omega_n\}$ which correspond to the
bound states of the charged massive field.

Defining a new $\psi$ function
\begin{equation}\label{Eq14}
\psi\equiv \Delta^{1/2}R\  ,
\end{equation}
the equation (\ref{Eq7}) can be written in
the form of a Schr\"odinger-like wave equation
\begin{equation}\label{Eq15}
{{d^2\psi}\over{dr^2}}+(\omega^2-V)\psi=0\  ,
\end{equation}
where
\begin{equation}\label{Eq16}
\omega^2-V={{U+M^2-Q^2}\over{\Delta^2}}\  .
\end{equation}

Then it was shown in \cite{Hod:2013eea,Hod:2013nn} that there is no superradiant amplification  for the charged Reissner-Nordstr\"om black holes because mainly two necessary conditions  cannot be satisfied
simultaneously namely, the condition of  superradiant amplification (\ref{rad2}) and the existence of a trapping potential well.

\section{Superrrandiance Instability of  the Reissner-Nordstr\"om black hole with the Derivative Coupling}
\label{dercoupl}

 In the previous Section we discussed the superradiance effect of a massive charge scalar
 field  coupled minimally to Gravity resulting from the action
\be \label{action2}
S_0=\int dx^{4} \sqrt{-g}\; \left[\fr{R}{16\pi G}-g_{\m\n}D^{\m}\Psi(D^{\n}\Psi)^*-\frac{1}{4}F^{\m\n}F_{\m\n}-\m^2|\Psi|^2\right]~,
\ee
 where $D_\m\equiv \nabla_\m-iqA_\m$ and $F_{\mu\nu} = \partial_\mu A_\nu - \partial_\nu A_\mu$. In what follows $G = c = 1$. We note here that the only scale in the theory is provided by the mass of the scalar field.

We wand to generalize this action adding a coupling of the scalar field to Einstein tensor
\be \label{action}
S_0=\int dx^{4} \sqrt{-g}\; \left[\fr{R}{16\pi G}-(g^{\m\n}-\beta G_{\m\n})D^{\m}\Psi(D^{\n}\Psi)^*-\frac{1}{4}F^{\m\n}F_{\m\n}-\m^2|\Psi|^2\right]~.
\ee
The  motivation for introducing this coupling is twofold. First of all, the coupling of the scalar field to the Einstein tensor $G_{\mu\nu}$ captures the strong gravity effects. As it was discussed in the introduction this coupling modifies the kinetic properties of the scalar field. The scalar field   feels the strong gravity force and the net result is its kinetic energy to be reduced \cite{germani,Koutsoumbas:2013boa,Koutsoumbas:2015ekk,Kuang:2016edj}. Therefore it would be interesting to see what is the effect of this coupling to the superradiance effect.

Varying the action (\ref{action}) we get the Einstein equations
\bea G_{\mu\nu} = 8\pi T_{\mu\nu} \
\ , \ \ \ \ T_{\mu\nu} = T_{\mu\nu}^{(\Psi)} +
T_{\mu\nu}^{(EM)} - \beta\Theta_{\mu\nu}~, \label{einst}
\eea
where $T_{\mu\nu}^{(\Psi)}$,
$T_{\mu\nu}^{(EM)}$  are the energy-momentum tensors of the scalar and electromagnetic fields
\bea
T_{\mu\nu}^{(\Psi)} & = &   \Psi_{\mu\nu} + \Psi_{\nu\mu} - g_{\mu\nu}(g^{ab}\Psi_{ab} + \m^2 |\Psi|^2)~, \\
T_{\mu\nu}^{(EM)} & = & F_{\mu}^{\phantom{\mu} \alpha} F_{\nu
\alpha} - \fr{1}{4} g_{\mu\nu} F_{\alpha\beta}F^{\alpha\beta}~,
\eea
while  $\Theta_{\mu\nu}$ is the contribution to the energy-momentum tensor of the $G_{\mu \nu}$ term
\bea
\Theta_{\mu\nu}  = & -& g_{\mu\nu} R^{ab}\Psi_{ab} + R_{\nu}^{\phantom{\nu}a}(\Psi_{\mu a} + \Psi_{a\mu}) + R_{\mu}^{\phantom{\mu}a} (\Psi_{a\nu} + \Psi_{\nu a})  - \fr{1}{2} R (\Psi_{\mu\nu} + \Psi_{\nu\mu}) \nn\\
& - & G_{\mu\nu}\Psi - \fr{1}{2}\nabla^a\nabla_\mu(\Psi_{a\nu} + \Psi_{\nu a}) - \fr{1}{2}\nabla^a\nabla_\nu(\Psi_{\mu a} + \Psi_{a\mu}) + \fr{1}{2}\Box (\Psi_{\mu\nu} + \Psi_{\mu\nu}) \nn \\
& + & \fr{1}{2}g_{\mu\nu} \nabla_a\nabla_b (\Psi^{ab} + \Psi^{ba})
+ \fr{1}{2}(\nabla_\mu\nabla_\nu + \nabla_\nu\nabla_\mu) \Psi -
g_{\mu\nu}\Box\Psi~. \label{theta} \eea
The Klein-Gordon equation is
\be
(\partial\mu-i q A_\mu) \left[
\sqrt{-g}(g^{\mu\nu} - \beta G^{\mu\nu})(\partial\nu - i q A_\nu)\Psi \right] =
\sqrt{-g} \m^2 \Psi~, \label{glgord}
\ee
and the Maxwell equations are
\be
\nabla_\nu F^{\mu\nu} +(g^{\mu\nu} - \beta G^{\mu\nu}) \left[ 2 q^2 A_\nu |\Psi|^2 + i q
(\Psi^*\nabla_\nu\Psi - \Psi\nabla_\nu\Psi^*)\right]=0~.
\label{max}
\ee
For convenience we had set
 \bea
\Psi_{\mu\nu} &\equiv& D_{\mu}\Psi (D_{\nu}\Psi)^*~,\\
\Psi &\equiv& g^{\mu\nu}\Psi_{\mu\nu}~.
 \eea

This action if $\beta=0 $ represents a massive charged scalar field with canonical kinetic terms coupled to gravity. If $\beta \neq 0 $ then
this scalar field is coupled directly to curvature through its coupling to the Einstein tensor $G_{\mu\nu}$. We have to solve the Einstein-Maxwell-scalar field equations (\ref{einst}), (\ref{glgord}) and (\ref{max}). To get an insight to the superradiance effect, we will first solve this system in the probe limit, assuming that the scalar field does not backreact to the metric. This means that the Reissner-Nordstr\"om metric solves the Einstein equations and the scalar field acts like a test field. In the next two sections we will study the superradiance effect of the backreacting solutions \cite{Kolyvaris:2011fk,Rinaldi:2012vy,Kolyvaris:2013zfa,Babichev:2013cya} resulting from the action (\ref{action})

Before we proceed we will discuss the motivation for considering the action (\ref{action}). As we discussed in the previous Section if we scatter a scalar field with canonical kinetic terms ($\beta=0$ in the action (\ref{action}) we considered) off the horizon of a Reissner-Nordstr\"om black hole, we do not see any superradiance amplification. The main reason for this is that because asymptotically the space is flat there is no any barrier or trapping potential to capture the reflected wave and inspite that the Bekenstein superradiance condition is satisfied we do not have an amplification of the scalar wave which would have to result in an instability of the Reissner-Nordstr\"om background metric \cite{Hod:2013eea,Hod:2013nn}. As we will discuss in the following, the presence of the derivative coupling modifies the Bekenstein superradiance condition and in the same time  generates  a confining potential outside the horizon of the Reissner-Nordstr\"om  black hole and this will lead to an instability of the background metric.

The action (\ref{action}) we considered  is part of the shift-symmetric Horndeski action \cite{Horndeski:1974wa}. The new information it introduces the derivative coupling $\beta$ of the scalar field to Einstein tensor, which has the dimensions of length squared,  is a  scale in the theory which effectively on short distances acts as a cosmological constant  \cite{Kolyvaris:2011fk,Rinaldi:2012vy,Kolyvaris:2013zfa,Babichev:2013cya}. Therefore, the space is not any more flat and as it was shown in \cite{Kolyvaris:2013zfa} a potential well is formed near the horizon of a Reissner-Nordstr\"om. Having this result, in this paper we will study the superradiance effect of a massive charge scalar field coupled to Curvature in the background of a  Reissner-Nordstr\"om black hole.

Considering the Reissner-Nordstr\"om  metric
\bea
ds^2 = -f(r)dt^2 + \fr{dr^2}{f(r)} + r^2 d\Omega^2~,
\eea
with $f(r) = 1 - 2M/r + Q^2/r^2$ and an electromagnetic potential $A_t = -\frac{Q}{r}$
we decompose the scalar field as in (\ref{Eq6})
\bea
\Psi = e^{-i\omega t}e^{i m \varphi}S(\theta)\frac{\psi(r)}{r}~. \label{wav11}
\eea
Then the radial equation resulting from the Klein-Gordon equation (\ref{glgord}) using (\ref{wav11}) can be written in a Schr\"odinger-like form
\bea\label{Schr}
\frac{d^2\psi}{dr_*^2}+U_{eff}(r)\psi(r_*)=0~,
\eea
where we have defined the ``tortoise" coordinate
\bea
\frac{dr_*}{dr} = \frac{r^2}{f(r) \left(r^2+\beta(1-f-rf')\right)}
\eea
and the effective potential is given by
\bea
U_{eff} &=& \left(\omega -\frac{q Q}{r}\right)^2-f \left[\frac{\ell(\ell+1)}{r^2}+\mu ^2+\frac{f'}{r^2}\right] \nonumber\\
&+&\beta  \left(-\frac{(f-1) \left(2 f^2-f \ell(\ell+1)+2 q^2 Q^2\right)}{r^4}+\frac{\left(f \mu ^2-2 \omega ^2\right) f'}{r}\right.\nonumber\\
&+&\left.\frac{4 (f-1) q Q \omega +2 \left(f \left[f-1+\ell(\ell+1)\right]-q^2 Q^2\right) f'}{r^3} \right.\nonumber\\
&+&\left.\frac{2 (f-1) \left(f \mu ^2-2 \omega ^2\right)+8 q Q \omega  f'+4 f \left(f'\right)^2+f \left[2 f+\ell(\ell+1)\right]
 f''}{2 r^2}\right)\nonumber\\
&+& \beta ^2 \left(\frac{(f-1)^2 \left(2 f^2+q^2 Q^2\right)}{r^6}+\frac{\omega ^2 \left(f'\right)^2}{r^2}\right.\nonumber\\
&+&\left.\frac{(f-1) \left(-2 (f-1) q Q \omega +\left(f (1+f-\ell(\ell+1))+2 q^2 Q^2\right) f'\right)}{r^5}\right.\nonumber\\
&-&\left.\frac{f' \left(-4 (f-1) \omega ^2+4 q Q \omega  f'+2 f \left(f'\right)^2+f \left(2 f+\ell(\ell+1)\right) f''\right)}{2 r^3}
\right.\nonumber\\
&+&\left.\frac{(f-1) \left(2 (f-1) \omega ^2-f \left(2 f+\ell(\ell+1)\right) f''\right)}{2 r^4}\right.\nonumber\\
&-&\left.\frac{8 (f-1) q Q \omega  f'+2 \left(f \left(2(f-1)+\ell(\ell+1)\right)-q^2 Q^2\right) \left(f'\right)^2}{2 r^4}\right)~. \label{Effpot}
\eea

Note that the effective potential because of the presence of the derivative coupling $\beta$ has the lapse function at second order and this  introduces high order terms in the radial coordinate $r$. Then the effective potential depends on seven parameters $U_{eff}(M, Q, \mu, q, \omega, l, \beta)$ and even its numerical study is difficult.

The effective potential at infinity goes like
\bea
U_\infty \sim \omega^2-\m^2~,
\eea
while near the horizon can be approximated as
\bea
U_{r_H} \sim \frac{(r_H^2+\beta \Phi^2)^2(\omega-q\Phi)^2}{r_H^4} +O(r-r_H)~,
\eea
where $\Phi=Q/r_H$ is the electric potential at the horizon.

In a scattering experiment the Klein-Gordon equation \eqref{Schr} has the following asymptotic behaviour
\bea
\psi\sim \left\{ \begin{array}{ll}
Te^{-i\sigma r_*} & \textrm{, as } r\rightarrow r_H\\
e^{-ir_*\sqrt{\omega^2-\m^2}} + Re^{ir_*\sqrt{\omega^2-\m^2}} & \textrm{, as } r\rightarrow\infty
\end{array} \right.
\eea
where we have set
\bea
\sigma \equiv \frac{(r_H^2+\beta \Phi^2)(\omega-q\Phi)}{r_H^2}~.
\eea

These boundary conditions correspond to an incident wave of unit amplitude from spatial infinity giving rise to a reflected wave of amplitude $R$ and a transmitted wave of amplitude $T$ at the horizon.

Since the effective potential is real, there exists another solution $\bar{\psi}$ to \eqref{Schr} which satisfies the complex conjugate boundary conditions \cite{Ching:1993gt}. The solutions $\psi$ and $\bar{\psi}$ are linearly independent and thus their Wronskian $W=\psi\frac{d}{dr_*}\bar{\psi}-\bar{\psi}\frac{d}{dr_*}\psi$ is independent of $r_*$.
Evaluating the Wronskian at the horizon and infinity respectively, we get
\bea
W(r\rightarrow r_H) &=& 2 i \sigma |T|^2, \\
W(r\rightarrow \infty) &=& -2 i (|R|^2-1)\sqrt{\omega^2-\mu^2}
\eea
and by equating the two values we get,
\bea
|R|^2 &=& 1 - \frac{\s}{\sqrt{\omega^2-\mu^2}}|T|^2~.
\eea
We can see that if $\s < 0$ the wave is superradiantly amplified, $|R|^2 > 1$ \cite{Benone:2015bst}. So, the superradiant   condition (\ref{rad2}) in the presence of the derivative coupling is modified to
\bea\label{supercond}
(r_H^2+\beta \Phi^2) (\omega - q \Phi) < 0~.
\eea

To see whether the superradiance will cause the instability of the Reissner-Nordstr\"om black hole, we need to check whether there exists a potential well outside the horizon to trap the reflected wave. If the potential well exists, the superradiant instability will occur and the wave will grow exponentially over time near the black hole to make the background Reissner-Nordstr\"om   black hole unstable.
We are interested in solutions of the radial equation \eqref{Schr} with the physical boundary conditions of purely ingoing waves
at the horizon and a decaying solution at spatial infinity. A bound state decaying exponentially at spatial infinity is characterized by $\omega^2 < \mu^2$. We choose the parameters $M, Q, \mu, q, \beta, \omega$ to meet this condition together with the superradiant condition \eqref{supercond}. In what follows we will set $\ell=0$ and fix the horizon at $r_H=1$.

\subsection{Case I: $\omega<q\Phi$ and $r_H^2+\beta\Phi^2>0$}

We will first examine the case where $\omega<q\Phi$ and $\beta>0$ so that \eqref{supercond} is satisfied. Our aim is to see if a trapping potential is formed outside the horizon of the black hole. Since the effective potential (\ref{Effpot}) is a high order polynomial in the radial coordinate $r$ and therefore its analytical treatment  is difficult, we will relay on its numerical investigation. We intend to study the effective potential (\ref{Effpot}) systematically varying its parameters to get mainly the following physical information. Firstly, what is the range of parameters $Q$ and $M$ for which a trapping potential is formed and secondly if the scalar field is strongly coupled to curvature (large $\beta$) does the potential well depends?  To do that we will fix the parameters $q$ and $\omega$,  $\mu$ as the relation $\omega^2 < \mu^2$ to be satisfied and leave the others free to vary.

\begin{figure}
\centering
\includegraphics[scale=1]{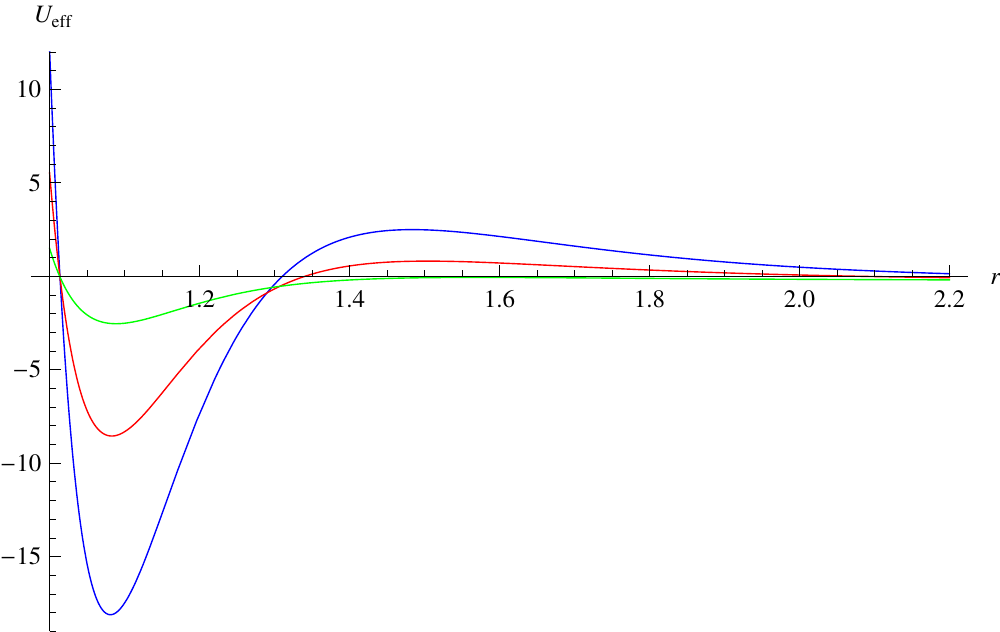}
\caption{The potential as a function of the radial coordinate for coupling constant $\beta=150,\ \beta=100,\ \beta=50$ (blue, red, green), mass and charge for the black hole and the scalar field $M=0.625,\ Q=0.5,\ \mu=0.63,\ q=0.86$ respectively and $\omega=0.34$.} \label{fig1a}
\end{figure}

\begin{figure}
\centering
\includegraphics[scale=1]{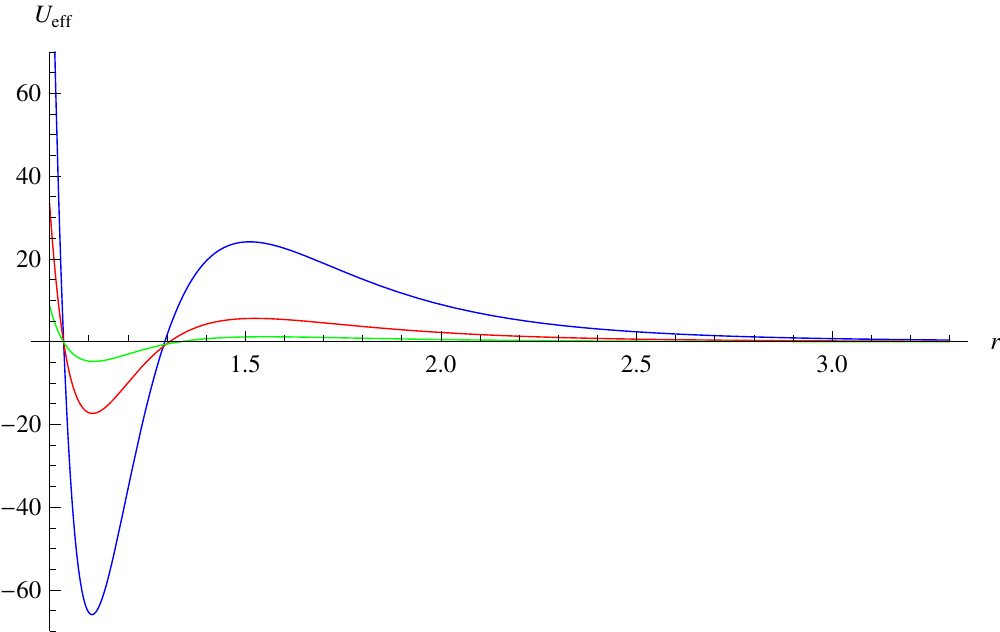}
\caption{The potential as a function of the radial coordinate for coupling constant $\beta=200,\ \beta=100,\ \beta=50$ (blue, red, green), mass and charge for the black hole and the scalar field $M=0.82,\ Q=0.8,\ \mu=0.63,\ q=0.86$ respectively and $\omega=0.6$.} \label{fig1b}
\end{figure}

In Table \ref{table1} we give the ratio $(Q/M)^2$ for various values of $Q$ and $M$.
Our criterion for building up the values that appear in  Table \ref{table1} was  that we looked among the various values of the ratio  $(Q/M)^2$ to such values of the ratio approaching one (near extremal case), one (extremal limit) and far way from one. For a characteristic value of $Q$ and $M$ in the first part of the Table \ref{table1} we depict the effective potential as a function of $r$ in  Figure \ref{fig1a}, while in Figure \ref{fig1b} we depict the effective potential as a function of $r$ for a characteristic value of $Q$ and $M$ in the second part of the Table \ref{table1}. We observe that irrespective of the value of the ratio $Q/M$ a trapped potential is formed outside the horizon of the black hole and most importantly  as the derivative coupling  $\beta$ is increased and the ratio $(Q/M)^2$ approaching the extremal limit the potential well gets deeper. This suggests that as the strength of the coupling of the scalar field to curvature is increased and the black hole approaches its extremal limit,  the superradiant instability is amplified.

\begin{figure}
\centering
\includegraphics[scale=0.95]{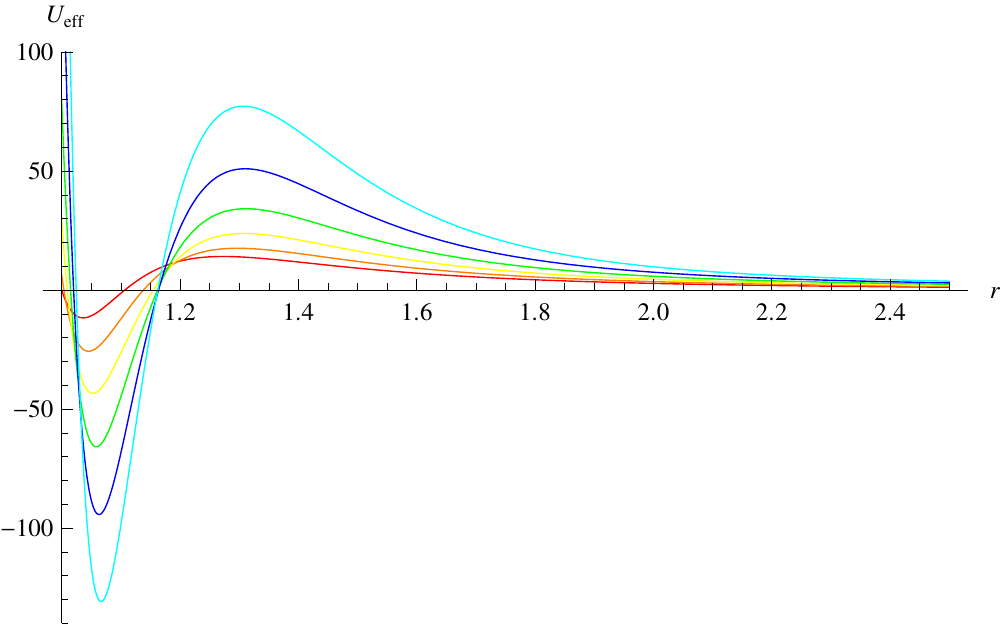}
\caption{The potential as a function of the radial coordinate for coupling constant $\beta=10$, mass and charge for the scalar field $q=0.86,\ \mu=1.63,\ \omega=1.28$. The ratio $(Q/M)^2$ is 0.852, 0.808, 0.760, 0.720, 0.679, 0.64 (red, orange, yellow, green, blue, cyan) respectively.} \label{fig1c}
\end{figure}

Finally in Figure \ref{fig1c} we depict the effective potential as a function of $r$ for fixed values of $\omega, \mu, q, \beta$ and for different values of $Q$ and $M$ belonging to the third part of Table \ref{table1}. We observe   that the potential well is deeper as the ratio $Q/M$ gets smaller. This corresponds to larger values of the mass $M$ and charge $Q$ since we are in the third part of  the Table \ref{table1}. This suggests that a highly charged massive black hole radiates more.


\begin{table}[h]
\begin{center}
\begin{tabular}{| c | c | c |}
\hline
\qquad M \qquad & \qquad Q \qquad &\qquad $(Q/M)^2$\qquad \\
\hline
 0.505 & 0.1 & 0.039 \\
 0.52 & 0.2 & 0.148 \\
 0.545 & 0.3 & 0.303 \\
 0.58 & 0.4 & 0.476 \\
 0.625 & 0.5 & 0.640 \\
 0.68 & 0.6 & 0.779 \\
 0.745 & 0.7 & 0.883 \\
 \hline
 0.82 & 0.8 & 0.952 \\
 0.905 & 0.9 & 0.989 \\
 1. & 1. & 1.000 \\
 1.105 & 1.1 & 0.991 \\
 1.22 & 1.2 & 0.967 \\
 1.345 & 1.3 & 0.934 \\
 1.48 & 1.4 & 0.895 \\
 \hline
 1.625 & 1.5 & 0.852 \\
 1.78 & 1.6 & 0.808 \\
 1.945 & 1.7 & 0.764 \\
 2.12 & 1.8 & 0.721 \\
 2.305 & 1.9 & 0.679 \\
 2.5 & 2. & 0.640\\
 \hline
\end{tabular}
\end{center}
\caption{The first and third sections correspond to a ratio less than 8/9, while the second greater than 8/9.}\label{table1}
\end{table}

We note that the behaviour shown in the above figures are characteristic of the non zero value of the derivative coupling. If $\beta=0$ then we can see from (\ref{Effpot}) that the effective potential does not develop a well and the we do not expect superradiance amplification  \cite{Hod:2013nn}.

\subsection{Case II: $\omega>q\Phi$ and $r_H^2+\beta \Phi^2<0$}

We consider now the case where $\omega>q\Phi$ and $\beta < - (r_H/\Phi)^2$ so again \eqref{supercond} is satisfied. This case is interesting.
 In \cite{Kolyvaris:2011fk, Kolyvaris:2013zfa}  fully backreacted black hole solutions were found in the presence of the derivative coupling $\beta$. However, these solutions exists only in the case of positive coupling while if the coupling constant $\beta$ is negative, then the system of Einstein-Maxwell-Klein-Gordon equations is unstable and no solutions were found.
  In \cite{Koutsoumbas:2013boa} a very small window of negative $\beta$ was shown to be allowed. For negative derivative coupling the stability of the Galileon black holes was investigated but there is no any conclusive result. For example in \cite{Minamitsuji:2014hha,Huang:2018kqr} the Black hole
quasinormal modes in a scalar-tensor theory with field derivative
coupling to the Einstein tensor were calculated.

In our study in this Section the scalar field  coupled to Einstein tensor does not backreact on the fixed Reissner-Nordstr\"om background metric. Therefore the physical propagating degrees of freedom do not interact with the background metric to alter it. The same is happening in Cosmology.  It is known that the effect of the derivative coupling in  a fixed FRW background metric, has the effect that for $\beta$ positive the Universe contracts
and if $\beta$  is negative the Universe expands \cite{Sushkov:2009hk}. The reason for this behaviour is  that the derivative coupling defines an effective cosmological constant. Also the stability of a static Universe was studied in \cite{Huang:2018kqr} where conditions for the coupling was found for both signs of of $\beta$. In the next Sections where we will study the backreacring effects we will take the derivative coupling to be positive (opposite sign to the usual  kinetic term).

\begin{figure}[h]
\centering
\includegraphics[scale=1]{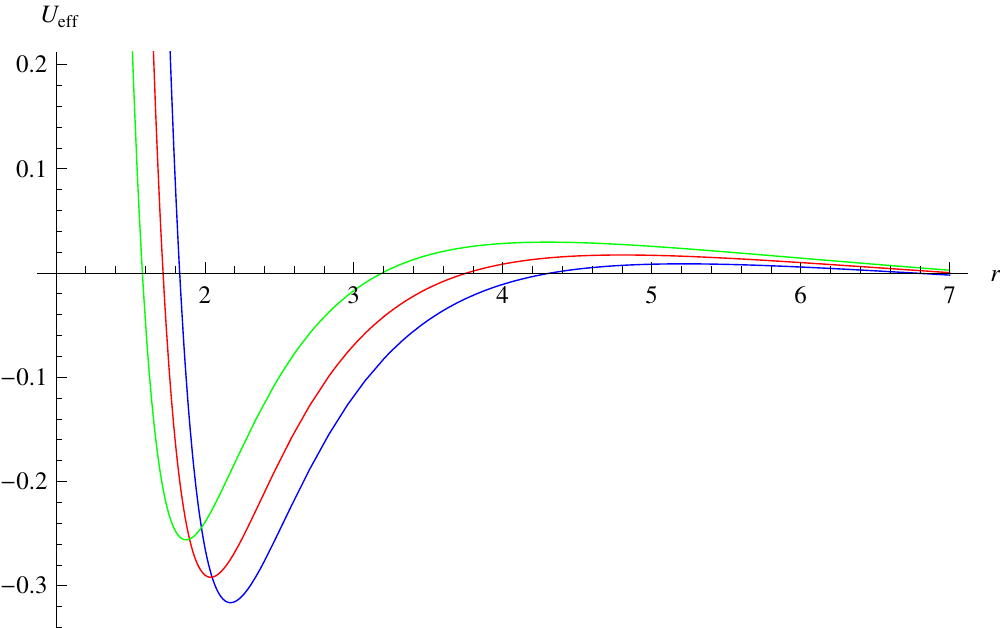}
\caption{The potential as a function of the radial coordinate for coupling constant $\beta=-45,\ \beta=-35,\ \beta=-25$ (blue, red, green), mass and charge for the black hole and the scalar field $M=0.625,\ Q=0.5,\ \mu=1.63,\ q=1.6$ respectively and $\omega=1.6$.} \label{fig2a}
\end{figure}

\begin{figure}[h]
\centering
\includegraphics[scale=1]{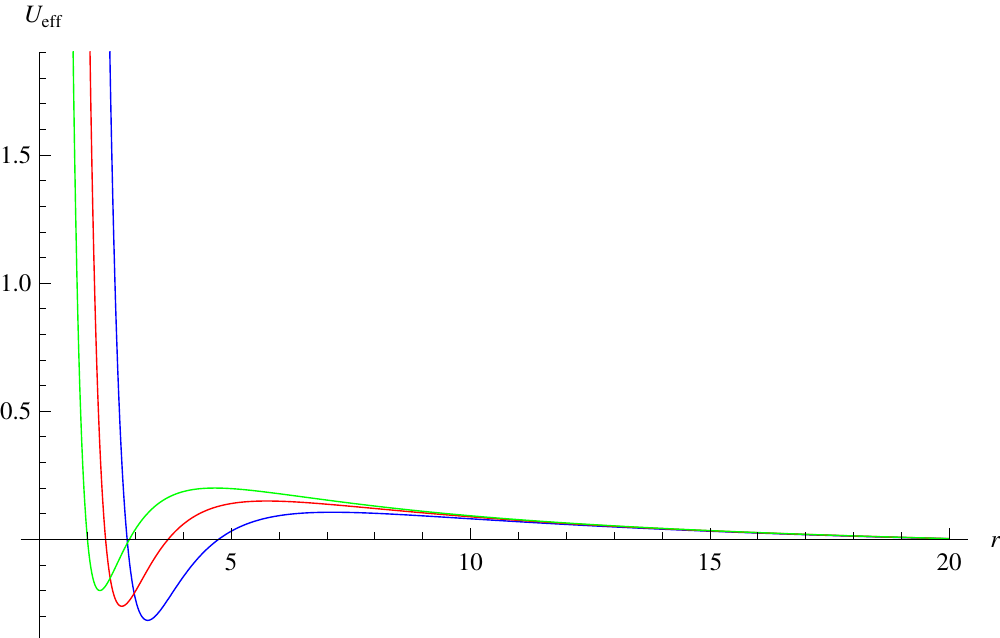}
\caption{The potential as a function of the radial coordinate for coupling constant $\beta=-100,\ \beta=-50,\ \beta=-25$ (blue, red, green), mass and charge for the black hole and the scalar field $M=0.82,\ Q=0.8,\ \mu=1.63,\ q=0.9$ respectively and $\omega=1.6$.} \label{fig2b}
\end{figure}

 In Figures \ref{fig2a} and \ref{fig2b} we depict the effective potential as a function of $r$, again for two characteristic values of the ratio $Q$ and $M$ belonging to the first and second part of Table \ref{table1}.  As before, in both cases we found that there exists a negative minima of the potential well which gets deeper for larger (absolute) values of the derivative coupling $\beta$. Therefore, for a range of parameters satisfying relation \eqref{supercond}, for $\omega>q\Phi$ and a negative $\beta$  there is a confining potential, indicating that  there should be an instability in the system of a charged scalar field scattered off a Reissner-Nordstr\"om black hole and this is due to superrandiant amplification.

\begin{figure}[h]
\centering
\includegraphics[scale=1]{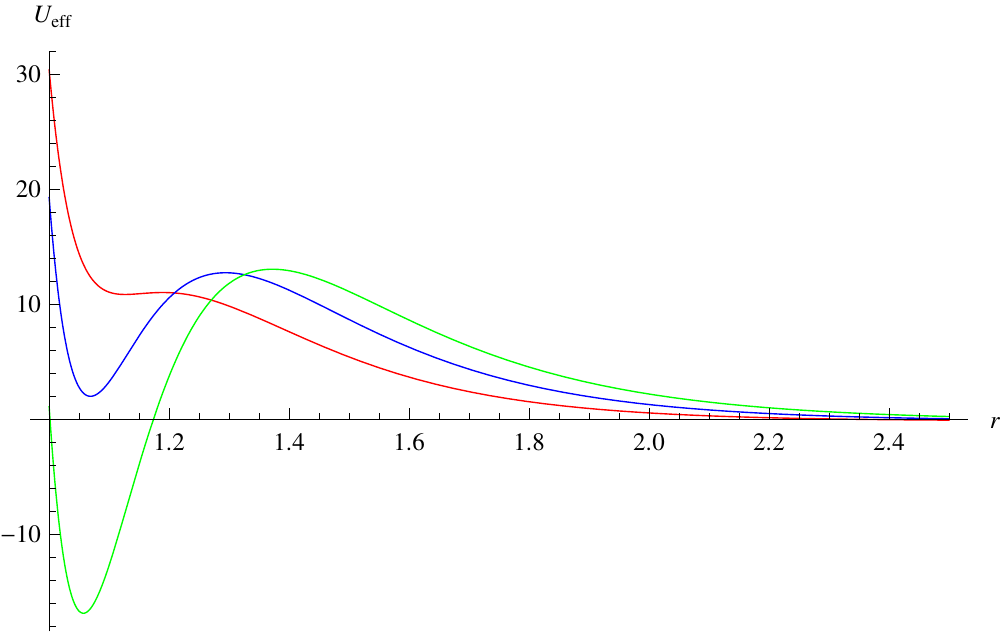}
\caption{The potential as a function of the radial coordinate for coupling constant $\beta=-120$, mass and charge for the scalar field $q=0.86,\ \mu=0.63,\ \omega=0.62$. The ratio $(Q/M)^2$ is 0.64, 0.779, 0.883 (red, blue, green) respectively.} \label{fig2c}
\end{figure}

\begin{figure}[h]
\centering
\includegraphics[scale=1]{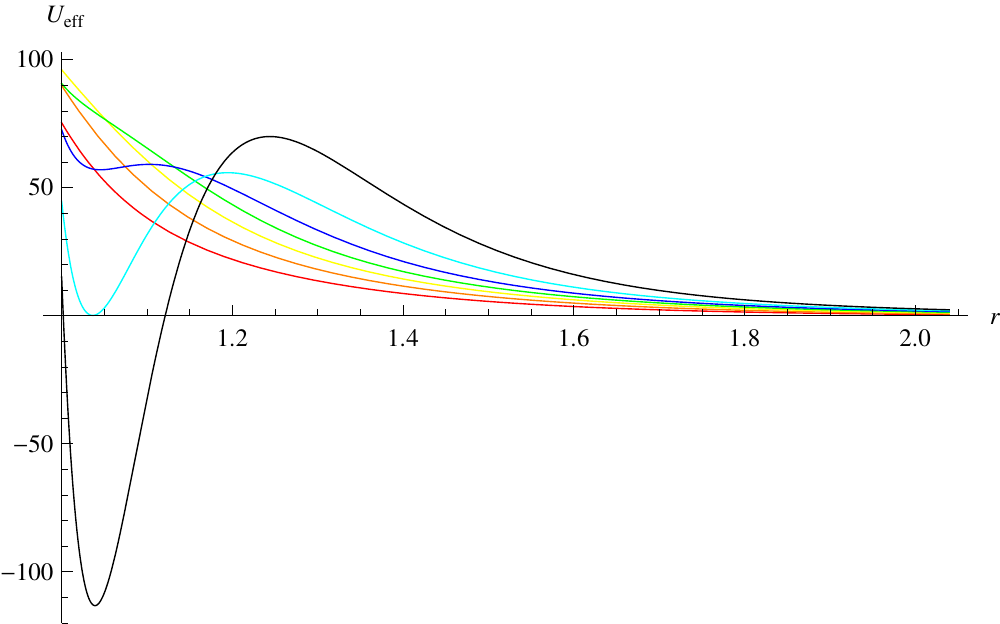}
\caption{The potential as a function of the radial coordinate for coupling constant $\beta=-50$, mass and charge for the scalar field $q=0.4,\ \mu=0.63,\ \omega=0.6$. The ratio $(Q/M)^2$ is 0.952, 0.989, 1.000, 0.991, 0.967, 0.939, 0.895 (red, orange, yellow, green, blue, cyan, black) respectively.} \label{fig2d}
\end{figure}

\begin{figure}[h]
\centering
\includegraphics[scale=1]{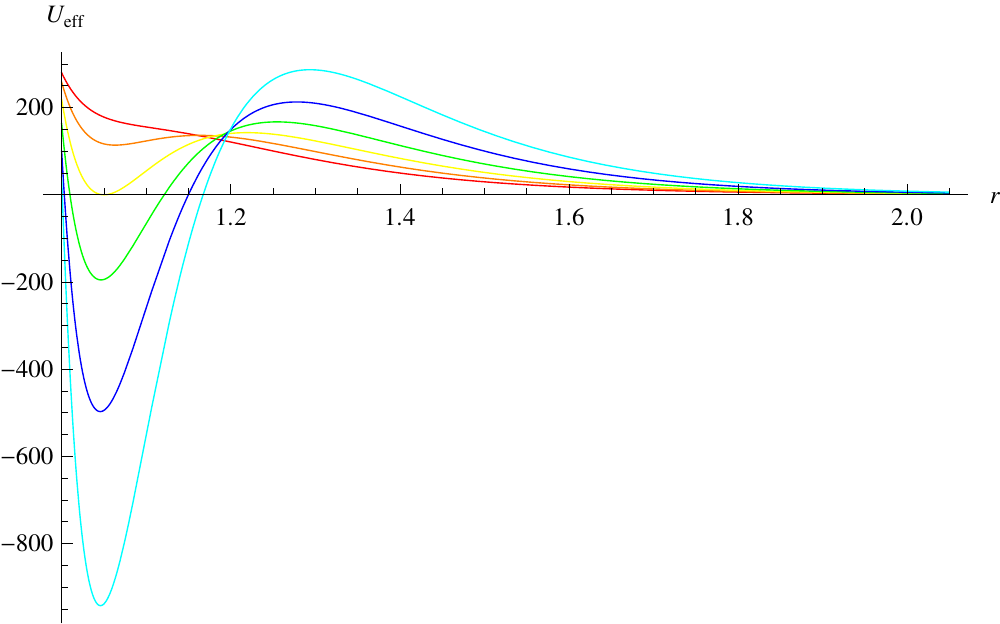}
\caption{The potential as a function of the radial coordinate for coupling constant $\beta=-20$, mass and charge for the scalar field $q=0.6,\ \mu=1.63,\ \omega=1.28$. The ratio $(Q/M)^2$ is 0.852, 0.808, 0.760, 0.720, 0.679, 0.64 (red, orange, yellow, green, blue, cyan) respectively.} \label{fig2e}
\end{figure}

An interesting behaviour is observed if we fix the value of the derivative coupling $\beta$. Then in Figure \ref{fig2c} we depict the effective potential as a function of $r$ for fixed values of $\omega, \mu, q, \beta$ and we vary the values of change $Q$ and mass $M$ belonging to the first part of Table \ref{table1}. We see that as the ratio is increased eventually a trapping potential is formed. Note that this happening as the charge $Q$ is increased.
In Figure \ref{fig2d} we depict the effective potential  for values of $Q$ and $M$ which  they belong to the second part of the Table \ref{table1}. We observe that a trapping potential is formed only for values of $Q$ and $M$ which they give a ratio that it is far away from the extremal limit or the near extremal limit. Finally in Figure \ref{fig2e} we see the formation of a trapping potential for values of $Q$ and $M$ belonging to the third part of Table \ref{table1}. We observe that as the mass $M$ and charge $Q$ in increased the potential well gets deeper.

\section{Superradiance in the Charged Galileon Black Hole with time dependent hair}

\label{Galileon1}

In the previous Sections we have investigated the instability of the Reissner-Nordstr\"om black hole, when the incident wave is described by a charged scalar field, coupled to gravity and to the Einstein tensor. In what follows, we will examine the case where the scalar field $\varphi$ coupled to  Einstein tensor backreacts  to the background metric. As shown in \cite{Babichev:2013cya}, a hairy black hole is generated, the charge  Galileon black hole. Then we will consider a complex, massive, charged scalar field $\phi$ of mass $\mu$ and charge $q$  scattered off the horizon of the Galileon black hole and we will study if superradiance radiation.

We consider a general action in which we have also included a coupling of the gauge field  to the scalar field $\varphi$
\bea \label{action1}
S= \int d^4 x \; \sqrt{-g} \; \Big{[} R-2\Lambda -\frac{1}{4} F^{\mu\nu} F_{\mu\nu} &-& \eta g_{\mu \nu}\partial
\varphi^{\mu} \partial \varphi^{\nu} +  \beta\,
G_{\mu\nu}\nabla^{\mu}\varphi\nabla^{\nu}\varphi
 -\gamma \,T_{\mu \nu}\,
\nabla^{\mu}\varphi\nabla^{\nu}\varphi \nonumber \\ &-&  \eta^\prime (D_\mu \phi)^* (D^\mu \phi)  - \mu^2  \phi \phi^* \Big{]},\nonumber \\
\eea
where $D_\mu = \nabla_\mu - i q A_\mu$ and $F_{\mu\nu} = \nabla_{\mu} A_\nu - \nabla_\nu A_\mu.$
Varying the action $\eqref{action1} $, we get the Einstein equations
\begin{eqnarray}
G_{\mu\nu}={T}_{\mu \nu}^{(1)}+{T}_{\mu \nu}^{(2)}+{T}_{\mu \nu}^{(3)}+{T}_{\mu \nu}^{(4)} +T_{\mu\nu}^{\phi}~,\label{eqmetric}
\end{eqnarray}
where we have defined,
\begin{eqnarray}\label{tmunu}
T_{\mu \nu}^{(1)}&=& -\Lambda\,g_{\mu \nu} + T_{\mu\nu}^{(M)}~,
\nonumber\\
T_{\mu\nu}^{(2)}  &
=&{\beta}\,\Big\{\frac{1}{2}\nabla_{\mu}\varphi\nabla_{\nu }\varphi
R-2\nabla_{\lambda}\varphi\nabla_{(\mu}\varphi
R_{\nu)}^{\lambda}-\nabla^{\lambda}\varphi\nabla^{\rho}\varphi
R_{\mu\lambda\nu\rho}
-(\nabla_{\mu}\nabla^{\lambda}\varphi)(\nabla_{\nu}\nabla_{\lambda}%
\varphi)+(\nabla_{\mu}\nabla_{\nu}\varphi)\square\varphi~,
\nonumber\\
&& +\frac{1}{2}G_{\mu\nu} (\nabla\varphi)^{2} -g_{\mu\nu}\left[
-\frac{1}{2}(\nabla^{\lambda}\nabla^{\rho}\varphi
)(\nabla_{\lambda}\nabla_{\rho}\varphi)+\frac{1}{2}(\square\varphi)^{2}%
-\nabla_{\lambda}\varphi\nabla_{\rho}\varphi R^{\lambda\rho}\right]\Big\}~,\nonumber \\
T_{\mu\nu}^{(3)} & = &\frac{1}{2}\,{\gamma}\,\left[F_{\mu\sigma}F_{\nu \rho}
\nabla^{\sigma }\varphi\,\nabla^{\rho }\varphi+\left(F_{\mu \sigma}
F^{\beta \sigma} \nabla_{\beta}\varphi \nabla_{\nu}\varphi+F_{\nu \sigma}
F^{\beta \sigma} \nabla_{\beta}\varphi \nabla_{\mu}\varphi\right)-\frac{1}{2}\, g_{\mu \nu}\,F_{\beta \sigma}F_{\tau}^{ \phantom{\tau} \sigma}\,\nabla^{\beta }\varphi\,\nabla^{\tau }\varphi\right.\nonumber\\
&&\left.+\frac{1}{8}\,g_{\mu \nu}\, \nabla^{\rho}\varphi\,
\nabla_{\rho}\varphi\, F_{\tau \beta} F^{\tau
\beta}-\frac{1}{2}\,F_{\mu \sigma} F_{\nu}^{\phantom{\nu}
\sigma}\,\nabla^{\rho}\varphi\,\nabla_{\rho}\varphi
-\frac{1}{4}\,\nabla_{\mu}\varphi\,\nabla_{\nu}\varphi\,F_{\tau \beta}
F^{\tau \beta}\right]~,\nonumber \\ T_{\mu\nu}^{(4)} & = & \eta\left(\partial_{\mu}\varphi\partial_{\nu}\varphi-\frac{1}{2} g_{\mu\nu}\partial_{\sigma}\varphi\partial^{\sigma}\varphi\right)~,\nonumber
\end{eqnarray}
with
\begin{eqnarray}\label{TE}
T_{\mu \nu}^{(M)}\equiv \frac{1}{2}\left( F_{\mu \sigma}
F_{\nu}^{\phantom{\nu} \sigma}-\frac{1}{4}\,g_{\mu \nu}\,F_{\alpha
\beta} F^{\alpha\beta}\right)~,
\end{eqnarray}
and
\begin{align*}
T_{\mu\nu}^{\phi} = -\frac{\mu^2}{2} g_{\mu\nu}{|\phi|^2} + \frac{\eta^\prime}{2} \Big(-g_{\mu\nu} \nabla_\kappa \phi^* \nabla^\kappa \phi + \nabla_\mu\phi^* \nabla_\nu \phi + \nabla_\mu\phi \nabla^\nu \phi^* - 2 q^2 A_\mu A_\nu |\phi|^2 \nonumber\\  - q^2 g_{\mu\nu} A_\kappa A^\kappa  |\phi|^2 - i q A^\kappa g_{\mu\nu}\left(\phi^* \nabla_\kappa \phi - \phi \nabla_\kappa \phi^* \right)
\Big)~.
\end{align*}
Assuming that the scalar field $\phi$ is a test field and it does not backreact  to the metric $($that is, keeping linear terms of $\phi)$ we end up with the coupled system of Einstein-Maxwell equations
\begin{align}
&G_{\mu\nu}={T}_{\mu \nu}^{(1)}+{T}_{\mu \nu}^{(2)}+{T}_{\mu \nu}^{(3)}+{T}_{\mu \nu}^{(4)}~, \\ &
\partial_{\mu}\left\{\sqrt{-g}\left[F^{\mu\nu}-\gamma\,\left(\frac{1}{2}\,F^{\mu \nu}\,\nabla_{\sigma}\varphi
+\big(F_{\sigma}^{\phantom{\mu} \mu}\nabla^{
\nu}\varphi-F_{\sigma}^{\phantom{\mu} \nu}\,\nabla^{\mu}\varphi
\big)\right)\,\nabla^{\sigma}\varphi\right]\right\}=0~,\label{eqmaxwell}
\end{align}
while the scalar field equation reads
\begin{equation}
\nabla_{\mu}\left[  \left(\beta \,
G^{\mu\nu}-\eta\,g^{\mu\nu}-\gamma\,T^{\mu \nu}_{(M)}\right)
\nabla_{\nu}\varphi\right]  =0~. \label{eqphi}
\end{equation}
A static spherically symmetric metric of the form,
\begin{equation}\label{ag}
    ds^{2}=-h(r)\,dt^{2}+\frac{dr^{2}}{f(r)}+r^{2}\big(d \theta^{2}+\sin^{2}\theta d \varphi^{2}\big)~,
\end{equation}
was assumed in \cite{Babichev:2015rva} and the scalar field was taken to be linearly time-dependent as
\begin{equation}\label{aphi}
    \varphi(t,r)=\mathfrak{q_s}\,t+ \psi(r)~,
\end{equation}
while the electromagnetic potential turned  out to be
  \begin{eqnarray}\label{aA}
    A_{\mu}dx^{\mu}=A(r)dt - P\cos(\theta)d\varphi~.
\end{eqnarray}
A solution to the above equations in the special case where $h(r) = f(r)$ was found
\begin{equation}\label{solfh}
\begin{aligned}
&h(r)= 1-{\frac {\mu}{r}} +{\frac {\eta\,{r}^{2}}{3\,\beta}}+{\frac { \gamma\, \left( {Q}^{2}+{P}^{2} \right) }{4\,\beta\,{r}^{2}}}~, \\
&(\psi'(r))^2 = \frac{1-f(r)}{f(r)^{2}} \mathfrak{q_s}^2~,\\
&F_{tr} =F(r)=\frac{Q}{r^{2}}, \quad F_{\theta \varphi}
=C(\theta)=P\,\sin(\theta)~.
\end{aligned}
\end{equation}
The coupling constants, along with the integration constants and the parameter  $\mathfrak{q_s}$ are related as
\begin{equation}\label{constfh}
{P}^{2}\beta\, \left( \Lambda\,\gamma+\eta \right) ={Q}^{2}\eta\,
 \left( \gamma-\beta \right)
~,\qquad \mathfrak{q_s}^2=\frac{\eta + \Lambda\,\beta}{\beta\,\eta}~,\qquad
C_{0}=\frac{1}{\eta} \left(\eta-\beta\,\Lambda\right)~.
\end{equation}

 Recently the stability of the  black hole solution (\ref{solfh}) was discussed in details in \cite{Babichev:2017lmw}. It was found that the couplings $\eta$ and $\beta$ should  have the opposite signs to guarantee the stability of the formed black hole solution.

\subsection{The dynamics of the scalar field $\phi$ outside the charge Galileon black hole}

We study now the  scattering of the test massive charged scalar field $\phi$ off the horizon of the charge Galileon black hole.  The dynamics is described by the Klein-Gordon equation $($we redefine $ \frac{\mu^2}{\eta^\prime} \equiv \mu_s^2 )$
\begin{equation}\label{kg}
\Big[ ( {\nabla}^{\nu} - iq A^{\nu} ) ({\nabla}_{\nu} -iq A_{\nu}) -{\mu}^2_{s} \Big] \phi = 0~.
\end{equation}
We  decompose  $\phi(t,r,\theta,\phi)$ as
\be
\phi(t,r,\theta,\phi) = e^{-i \omega t} R(r) Y(\theta,\phi)~. \label{wave1}
\ee
In the above decomposition $R$ and $Y$ denotes the
radial and angular part of the solution. Then the Klein-Gordon equation takes the form
\begin{equation}
\begin{aligned}
&- \frac{1}{f(r)} \frac{{\partial}^2 \phi}{\partial t^2} + f \frac{{\partial}^2 \phi}{\partial r^2}+ \frac{1}{r^2}\frac{{\partial}^2 \phi}{\partial {\theta}^2} +\frac{1}{r^2 \sin^2 \theta} \frac{{\partial}^2 \phi}{\partial \phi^2} +\frac{\cos \theta}{r^2 \sin \theta} \frac{\partial \phi}{\partial \theta} + \frac{2 i q P}{r^2} \frac{cos{\theta}}{sin^2 {\theta}} \frac{\partial \phi}{{\partial}_{\phi}} - \frac{q^2 P^2}{r^2} \frac{cos^2 {\theta}}{sin^2 {\theta}} \phi  \\
&+ \Big( \frac{r f'(r)+2f}{r} \Big) \frac{\partial \phi}{\partial r}  - \frac{2iqQ}{r f (r)} \frac{\partial \phi}{{\partial}_t}+ \Big( \frac{q^2 Q^2 - {\mu}^2_s r^2 f(r)}{r^2f(r)} \Big) \phi  = 0~,
\end{aligned}
\end{equation}
while  the angular part takes the form
\begin{equation} \label{jacobi}
\begin{aligned}
\frac{1}{r^2}\Bigg[ \frac{{\partial}^2 }{\partial {\theta}^2} - \frac{ m^2}{ \sin^2 \theta}  +  \frac{\cos \theta}{\sin \theta} \frac{\partial }{\partial \theta} + \frac{2 i^2 m q P cos{\theta}}{sin^2 {\theta}}  - \frac{q^2 P^2 cos^2 {\theta}}{sin^2 {\theta}}  \Bigg] Y(\theta) = \lambda Y(\theta)~.
\end{aligned}
\end{equation}
Applying the transformations $Y \rightarrow \frac{1}{\sqrt{\sin \theta}} \tilde{Y}$ this equation can be written in the form of the Jacobi differential equation, and  it admits the corresponding solution with eigenvalues
\begin{equation}
 \lambda = -(n+1) ( n+qP)~.
\end{equation}
 When the magnetic charge $P$ is zero, the Jacobi equation reduces to Legendre differential equation, with eigenvalues $n(n+1)$, while the eigenfunctions turn out to be the spherical harmonics $Y_{l m}$. Meanwhile, when $P = 0 $ and $\beta=0$ the metric reduces to the Reissner-Nordstr\"om spacetime, as discussed in Section II.

Now we can go back to Klein-Gordon of the full solution, which, after substitutions, reduces to the radial Klein-Gordon equation
\begin{equation}
f(r) \frac{d^2R(r)}{dr^2} + \Big( \frac{rf'(r)+2f(r)}{r} \Big) \frac{dR(r)}{dr}  +\Bigg[ \frac{ (\omega r-qQ)^2 - \Big((n+1)(n+qP) +{\mu}^2_s r^2 \Big)f(r)}{r^2 f(r)} \Bigg] R(r) = 0~.
\end{equation}
When $r \rightarrow r_H $ we have  $\quad f(r) = 0 $ and then the third term  is divergent at the horizon. To avoid this, we take the  coordinate transformation to the tortoise coordinate $r^*$, $dr^*= \frac{1}{f(r)}dr$. Then the radial Klein-Gordon equation becomes
\begin{equation}
R''(r^*)+\frac{2 f(r) R'(r^*)}{r} + \Bigg[ \frac{ ( \omega r - qQ )^2 - \left( (n+1)(n+qP)+\mu^2_s r^2\right)  f(r)}{r^2} \Bigg] R(r^*) = 0~.
\end{equation}
 We can bring this equation in a Schr{\"o}dinger-like form
 \begin{equation}
\frac{d^2R(r^*)}{dr^{*2}}+ \frac{U_{eff}(r)}{r^2} R(r^*) = 0~,
\end{equation}
where the $U_{eff}$ is the effective potential, given by the relation
\begin{equation}
U_{eff}(r) = ( \omega r - qQ)^2 -  f(r)\left[(n+1)(n+qP)+r f'(r)+m^2 r^2\right]~.
\end{equation}
As we can see, the effective potential encodes all the information about the background and the test scalar field. More specifically, the $f(r)$ metric function has the derivative coupling constant $\beta$ as a parameter that interest us.


Now we have to determine the solutions of the radial component of the scalar field at the boundaries. Let us assume that the potential is rea1. Then, since the background is stationary, the field equations are invariant under the transformations $t \rightarrow -t$ and $\omega \rightarrow -\omega$. Thus, there exists another solution $\bar{R}$ which satisfies the complex conjugate boundary conditions. The solutions $R$ and $\bar{R}$	are linearly independent and thus  their Wronskian is independent of $r^*$.

At the horizon, $f(r)=0$ and $U_{eff} (r) = (r \omega - qQ )^2$, so we get the solution
\begin{equation}
R(r^*) = C_{1,h} e^{i\frac{(r_h \omega -qQ ) r^*}{r_h}} + C_{2,h}e^{-i\frac{(r_h \omega -qQ ) r^*}{r_h}}~,
\end{equation}
where the presence of a horizon compels the amplitude $C_{1,h}$ to be zero, while $C_{2,h}$ is the amplitude of the transmitted wave.

To study the equation's behaviour at large r, we first have to notice that in this case, the fact that our metric is not asymptotically flat, affects the form of the effective potential which at large distances is,
\begin{equation}
U_{eff} = \omega^2 - \mu^2_s - \frac{2 \eta}{3 \beta} - \frac{\eta(n+1)(n+qP)}{3\beta}- r^2 \Big(\frac{\mu^2_s \eta}{3\beta} + \frac{2 \eta^2}{9 \beta^2} \Big)~.
\end{equation}

If we want to have a Schrodinger-like equation for the radial component of the scalar field at the $r \rightarrow \infty$ limit, we have to impose the following relation between the coupling constants
\begin{equation}
 \eta = - \frac{3 \mu^2 \beta}{2 \eta^\prime}~. \label{const1}
\end{equation}
Then $U_{eff}$ equals now a constant and the resulting solution reads:
\begin{equation}
R(r^*) = C_{1,\infty} e^{i \; \Big(\sqrt{ \omega^2 - \mu^2_s - \frac{2 \eta}{3 \beta} - \frac{\eta(n+1)(n+qP)}{3\beta}} \Big)r^*} + C_{2,\infty} e^{-i \; \Big(\sqrt{ \omega^2 - \mu^2_s - \frac{2 \eta}{3 \beta} - \frac{\eta(n+1)(n+qP)}{3\beta}} \Big)r^*}
\end{equation}
where $ C_{2,\infty}$ corresponds to the amplitude of an incident wave from spatial infinity and $C_{1,\infty}$ to the amplitude of the reflected wave.

\subsubsection{Superradiance Condition}

 Now that we have determined the solutions of the $R$ component of the scalar field at the boundaries, we are able to evaluate the Wronkians of two linearly independent solutions at the boundaries
\begin{equation}
\begin{aligned}
&W(r \rightarrow r_H) = -2i k_H |C_{2,H}|^2~,\\
&W(r \rightarrow \infty )= 2i k_{\infty} (|C_{1,\infty}|^2 - |C_{2,\infty}|^2)~.
\end{aligned}
\end{equation}
Taking the two expressions to be equal, we find
\begin{equation}
|C_{1,\infty}|^2 =  |C_{2,\infty}|^2 - \frac{k_H}{k_{\infty}}|C_{2,H}|^2~.
\end{equation}
For $\frac{k_h}{k_{\infty}} < 0$ the wave is superradiantly amplified.
In this case, $k_H =\frac{ \omega r_H - qQ}{r_H}$ and $k_{\infty} =\sqrt{\omega^2 - \mu^2_s - \frac{2 \eta}{3 \beta} - \frac{\eta(n+1)(n+qP)}{3\beta}}$, so we end up with the inequality
\begin{equation}
\frac{(r_H \omega -qQ)}{\sqrt{\omega^2 - \mu^2_s - \frac{2 \eta}{3 \beta} - \frac{\eta(n+1)(n+qP)}{3\beta}}} < 0~,
\end{equation}
which leads us to the superradiant condition
\begin{equation}
r_H \omega \;< qQ\;~.
\end{equation}
This condition is the Bekenstein superradiance condition for the Galileon black hole with a time dependent scalar field backreacting to the metric. The information of the derivative coupling enters the final inequality via the $r_H$, affecting the superradiance condition.
We also have to underline the fact that, only waves with
\begin{equation}
\omega^2 > \mu^2_s + \frac{2 \eta}{3 \beta} + \frac{\eta(n+1)(n+qP)}{3\beta}~,
\end{equation}
propagate to infinity restricting in this way the allowed values of the frequency $\omega$ of the incitant wave.

Before ending this Section we note that the constraint  (\ref{const1}) which it is imposed in order to have superradiance, connects the scattered test wave parameters of its mass and its kinetic energy with the parameters of the background Galileon black hole.

\section{Superradiance in the Charged Galileon Black Hole with  static hair}
\label{Galileon2}

We will study the superradiance effect of a specific model of a hairy Galileon black hole \cite{Cisterna:2014nua}. In this model the scalar field is
coupled to the background only with the Einstein tensor. Under the presence of an electric field, a
asymptotically locally flat hairy black hole solution was obtained  in the presence of a
cosmological constant. The action is given by
\begin{equation}
I[g_{\mu\nu},\varphi]=\int\sqrt{-g}d^{4}x\left[  \kappa\left(  R-2\Lambda
\right)  +\frac{\beta}{2}G_{\mu\nu}\nabla^{\mu}\varphi\nabla^{\nu}\phi-\frac{1}%
{4}F_{\mu\nu}F^{\mu\nu}\right]  \ ,
\end{equation}
where $\kappa=\frac{1}{16 \pi G}.$ Then the following solution was found
\begin{equation}
ds^{2}=-F(r)dt^{2}+\frac{15[4\kappa r^{2}(2-\Lambda r^{2})-q^{2}]^{2}}{r^{4}%
}\frac{dr^{2}}{F(r)}+r^{2}d\Omega^{2}\ \label{solu2},
\end{equation}
where
\begin{align}
F(r)  &  =48\kappa^{2}\Lambda^{2}r^{4}-320\kappa^{2}\Lambda r^{2}%
+120\kappa(8\kappa+\Lambda q^{2})-\frac{\mu}{r}+240\kappa\frac{q^{2}}{r^{2}%
}-5\frac{q^{4}}{r^{4}}\ ,\label{metfun}\\
\psi(r)^{2}  &  =-\frac{15}{2}\frac{(4\kappa\Lambda r^{4}+q^{2})(4\kappa
r^{2}(2-\Lambda r^{2})-q^{2})^{2}}{r^{6}\beta}\frac{1}{F(r)}\ ,\\
A_{0}(r)  &  =\sqrt{15}\left(  \frac{q^{3}}{3r^{3}}-8\kappa\frac{q}{r}%
-4\kappa\Lambda rq\right)\label{poten}  \ ,
\end{align}
where $\psi(r)=\varphi'(r)$. Observe that the derivative coupling $\beta$ does not appear in the metric function but only in the scalar field solution.

As before we consider a massive test scalar wave  $Y=Y_{lm}(t,r,\theta,\phi)$ in the background of the above black hole solution. The dynamics of this wave is described by the Klein-Gordon equation
\begin{equation}
[(\nabla^\nu -iqA^\nu )(\nabla_\nu -iqA_\nu)-\mu^2]Y=0~,
\end{equation}
where electromagnetic  potential $A_\nu$ is given by (\ref{poten}). Then writing the scalar field as
\begin{equation}
Y_{lm} = e^{im\phi} S(\theta)R(r)e^{-i\omega t}~,
\end{equation}
where $\omega$ is the conserved frequency, $l$ is the spherical harmonic, m is the azimuthal harmonic. R and S denote the radial and angular part,
 the radial Klein-Gordon equation is
\begin{equation}
\Delta\bigg{(} \frac{d}{dr}\Delta \frac{dR}{dr} \bigg{)}+UR=0~, \label{Eq. delta}
\end{equation}
where
\begin{equation}
\Delta=\frac{Fr^2}{\sqrt{X}}~,
\end{equation}
and
\begin{equation}
U=\big{(}\omega r^2 - qAr^2\big{)}^2-\Delta\big{(}\sqrt{X}l(l+1)+\mu^2 \sqrt{X}r^2\big{)}~.
\end{equation}
For our convenience we have define $X=\dfrac{15[4\kappa r^2(2-\Lambda r^2)-Q^2]^2}{r^4}$.
Defining a new function $\Psi$
\begin{equation}
\Psi=\Delta^{\frac{1}{2}}R~, \label{Eq. like shr.}
\end{equation}
the equation $(\ref{Eq. like shr.})$ can be written in the form of a  Schrondinger-like wave equation
\begin{equation}
\frac{d\Psi ^2}{dr^2}+V\Psi =0~,
\end{equation}
where
\begin{equation}
V=\frac{1}{4\Delta^2}\bigg{(}\big{(}\frac{d\Delta}{dr}\big{)}^2-2\Delta\frac{d^2\Delta}{dr^2}+4U\bigg{)}~.
\end{equation}
To solve the radial Klein-Gordon equation (\ref{Eq. delta}) we define the "tortoise" coordinate
\begin{equation}
\frac{dr_*}{dr}=\dfrac{\sqrt {X(r)}}{F(r)}~.
\end{equation}
Then the equation (\ref{Eq. delta}) takes the form
\begin{equation}
\frac{d^2 R}{dr_* ^2}+\frac{U}{r^4}R=0~.
\end{equation}

\subsubsection{Superradiance  Condition }

We impose boundary conditions of purely ingoing waves at the black-hole horizon
and a decaying solution at spatial infinity. We calculate the frequencies
 \begin{align}
 &k_{H}^2=V_{eff}(r\rightarrow r_+)=( \omega- qA(r_H))^2~,\\
 &k_{\infty}^2=V_{eff}(r\rightarrow \infty)=\infty~.
 \end{align}
At spatial infinity $k_{\infty}^2 \rightarrow \infty$, so there is no solution for the wave at infinity.

To overcome this problem we have to impose conditions on the parameters as in the case of the Galileon black hole we studied in Section IV. The  easiest choise is to put  the cosmological constant $\Lambda = 0$. Then the metric (\ref{solu2})  takes the form
\begin{equation}
ds^2 = -F(r)dt^2 + \dfrac{3(8\kappa r^2 - Q^2)^2}{r^4}\dfrac{dr^2}{F(r)}+r^2 d\Omega ^2~,
\end{equation}
where $F(r)$ and $X(r)$ are
\begin{eqnarray}
&&F(r)=192\kappa ^2-\dfrac{\mu}{r}+48\kappa \dfrac{Q^2}{r^2} - \dfrac{Q^4}{r^4}~, \\
&&X(r)=\dfrac{3(8\kappa r^2 -Q^2)^2}{r^4}~.
\end{eqnarray}
The electromagnetic potential of the black hole is
\begin{equation}
A_0 (r)=\sqrt{15}\big{(}\frac{Q^3}{3r^3} - 8\kappa \frac{Q}{r}\big{)}~,
\end{equation}
and it goes to zero at infinity. as expected.

Following the same procedure, with the new functions $X, F, A_0$, we solve the equation the radial equation (\ref{Eq. delta}) and take purely ingoing waves at the black hole horizon and a decaying solution at spatial infinity
\begin{equation}
R\sim \exp{-i\big{(}\omega- qA(r_+) \big{)}r_*}, \qquad  r\rightarrow r_+ ~ (r_* \rightarrow -\infty)~,
\end{equation}
and
\begin{equation}
R\sim  \exp\bigg{\{} -i\sqrt{\omega ^2 -192\kappa  ^2 \mu^2} r_*\bigg{\}}  , \qquad r\rightarrow \infty ~ (r_* \rightarrow \infty )~.
\end{equation}
Evaluating the Wronskian at the horizon and infinity respectively, and equating the two values we get
\begin{equation}
|R| ^2 = | I | ^2  -\dfrac{\omega- qA(r_+)}{\sqrt{\omega ^2 - 192 \kappa ^2 \mu ^2}}(| T| ^2)~.
\end{equation}
We can see that if
\begin{equation}
\omega < q\sqrt{15}\big{(}\frac{Q^3}{3r_H^3} - 8\kappa \frac{Q}{r_H}\big{)}~,
\end{equation}
the wave is superradiantly amplified.

\section{Conclusions}
\label{conclusion}

In this work we studied the superradiant effect of a class of  scalar-tensor Horndeski theory. We considered first a massive charge scalar field coupled to Einstein tensor with a derivative coupling, scattered off the horizon of  a  Reissner-Nordstr\"om black hole.  We  showed
that this
derivative coupling  provides a scale for a confining potential, and in the same time modifies the Bekenstein's superradiance condition (\ref{rad2}) with the derivative coupling appearing explicitly in the superradiance condition. We found  that for a wide range of parameters the superradiant condition is satisfied and in the same time a trapping potential is formed outside the horizon of a Reissner-Nordstr\"om black hole. As the strength of the coupling of the scalar field to curvature is increasing the depth of the potential is increasing indicating the amplification of the superradiant instability. The same amplification occurs for constant coupling as the mass and charge of the black hole is increasing.

This superradiance instability indicates that the background Reissner-Nordstr\"om black hole may require scalar hair. Then we studied the  backreacted effect. We  allowed the scalar field coupled to Einstein tensor to backreact to a charged spherical symmetric background. This leads to the generation of hairy charge Galileon black holes. The basic property of these solutions is that the derivative coupling appears as a parameter in these hairy solutions. We  studied  the superradiance effect and found the superradiance conditions of a massive charged scalar wave scattered off the horizon of these Galileon black holes. We  discussed two specific solutions. The first solution is a charge Galileon black hole solution with a time dependent scalar hair   and in the second solution the scalar hair is static.

In both cases the derivative coupling, which shows how strong the scalar field is coupled to curvature, influences  the superradiat radiation. In the case of a fixed Reissner-Nordstr\"om black hole background, the recover the Bekenstein superradiance condition for any value of the derivative coupling $\beta $ if $\beta> 0 $, while if $\beta < 0 $ then the Bekenstein superradiance condition is modified. In the case that the massive charge scalar field is scuttered off the horizon of the charge Galileon black hole the derivative coupling appears explicitly in the superradiance condition the strength of which influences the superradiance radiation in the case  the scalar hair of the Galileon black hole is time-dependent, while the superradiance condition is independent of $\beta$ if the scalar hair is static for the particular solution we considered.

\acknowledgments

We thank Vitor Cardoso for his valuable comments and remarks. The work of T.K. was funded by the FONDECYT Grant No. 3140261.


\end{document}